\documentstyle[preprint,aps]{revtex}
\newcommand{\be}{\begin{equation}} \newcommand{\ee}{\end{equation}} 
\newcommand{\bea}{\begin{eqnarray}}\newcommand{\eea}{\end{eqnarray}}

\begin{document}
\draft
\preprint{MRI-PHY/96-26, cond-mat/9610024}
\title{Calogero-Sutherland Type Models in Higher Dimensions}
\author{Pijush K. Ghosh$^{*}$}
\address{The Mehta Research Institute of
Mathematics \& Mathematical Physics,\\
Allahabad-211002, INDIA.}
\footnotetext {$\mbox{}^*$ E-mail: 
pijush@mri.ernet.in }  
\maketitle
\begin{abstract} 
We construct two different Calogero-Sutherland type models with
only two-body interactions in arbitrary dimensions. We obtain some
exact wave functions, including the ground states, of these two
models for arbitrary number of spinless nonrelativistic particles.
\end{abstract}
\narrowtext

\newpage

There are many exactly solvable models in one dimension,
described by $N$ fermions confined in a harmonic oscillator
potential or constrained to move on the rim of a
circle and interacting through each other via certain
special type of two-body interactions \cite{cs,cs1,pr,hs,ap}.
These models, namely the Calogero-Sutherland Model (CSM) and
its variants, have received a renewed interest in the recent
literature, because of their relevance in diverse branches of
physics \cite{chart}. Unfortunately, nothing much is known in
higher dimensions. This is primarily because of two reasons.
Firstly, there are not many known higher dimensional models,
where the fermionic description is possible for arbitrary
number of particles. Secondly, the inevitable appearance of
three-body interactions in the known higher dimensional analogue
of CSM makes any further analysis highly
nontrivial \cite{cs2,laugh,mbs,pijush,at}.

The purpose of this letter is to construct higher dimensional many-body
Hamiltonians with only two-body interactions, where the fermionic
description of the wave functions is possible for arbitrary number of
particles. With this purpose in mind, we first look for the possible
ways one can construct many-body fermionic wave functions, i.e.,
the wave functions which are antisymmetric under the exchange of any two
particle coordinates. This is also equivalent of looking for possible
correlations in many-body fermionic systems. We then identify an unique
correlation in arbitrary dimensions $D$ which can be realized by model
systems with only two-body interactions. The correlation
is such that the wave functions vanish whenever two particles
are on the surface of a $S^{D-1}$ whose center coincides with the origin
of the coordinate system. We construct two different models with only
two-body interactions which realize such correlations. We
obtain a set of exact solutions of these models. The set of exact solutions
include only the bosonic ground state and some excited states for both bosons
as well as fermions.

The exact wave functions of the CSM are highly correlated. These correlations
are encoded in the wave functions in the form of a Jastrow factor. This
Jastrow factor also takes care of the fermionic/bosonic nature of the
wave functions. In particular,  the Jastrow factor of the CSM has the form,
\be
J(x_1, x_2, \dots, x_N)=\prod_{i < j} X_{ij}^\lambda
{\mid X_{ij} \mid}^\alpha, \ \ X_{ij}=x_i - x_j,
\label{eq0}
\ee
\noindent where $x_i$ denotes the
particle position of the $i$th particle. The parameters $\alpha$ and
$\lambda$ are related to the
strength of the inverse square two-body potential. The Jastrow factor
$J$ has two interesting properties :\\
(a) $J$ vanishes when the position vectors
of any two particles coincide. This ensures that no two particles can occupy
the same position at the same time. Also, the two-body interaction in the
Hamiltonian
has singularities, precisely at these points, i.e., at $x_i = x_j$.\\
(b) $J$ picks up a factor $(-1)^\lambda$
under the exchange of particle indices. Consequently, spinless bosons as
well as fermions can be described by putting $\lambda$ equal to zero
or one, respectively.\\   
These two properties will be our basic criteria in constructing higher
dimensional analogue of (\ref{eq0}). The only freedom we have is the
modification of the property (a) to accommodate more zeroes in $J$, other
than those corresponding to the coinciding position vectors \cite{mbs,pijush}.
We would expect that the zeroes of $J$ will show up as the singularities
of the many-body Hamiltonian which realizes such correlations. In particular,
the two-body and the three-body interactions in the corresponding
Hamiltonian have the forms,
\bea
& & W_2 = \frac{g}{2} \sum_{i \neq j}  \left [ \vec{\bigtriangledown}_i .
\vec{\phi}(\vec{r}_i, \vec{r}_j) + g \vec{\phi}(\vec{r}_i,
\vec{r}_j) . \vec{\phi}(\vec{r}_i, \vec{r}_j) \right ],\nonumber \\
& & W_3 =  \frac{g^2}{2} \sum_{i \neq j \neq k}
\vec{\phi}(\vec{r}_i, \vec{r}_j) . \vec{\phi}(\vec{r}_i, \vec{r}_k), \ \
\vec{\phi}(\vec{r}_i, \vec{r}_j) = 
\frac{\vec{\bigtriangledown}_i X_{ij}}{X_{ij}},
\label{eq0.1}
\eea
\noindent where $g= \alpha + \lambda$. The above forms of $W_2$ and
$W_3$ can be obtained from the kinetic energy operator of the
Hamiltonian, i.e. $- \frac{1}{2} \sum_i \bigtriangledown_i^2 J$. $W_3$
vanishes for the Calogero model \cite{cs}. We will construct
below higher dimensional models where $W_3=0$. 

Note that $X_{ij}$ in (\ref{eq0}) is made out of the position
vectors of two particles. This is because the many-body
interactions encountered in physical systems can be factorised
in terms of
two-body interactions only. Moreover, the fermionic description
for arbitrary number of particles is not possible for $X$ having more
than two indices even if it satisfies the basic criteria.
To see this, let us consider a (pseudo-)scalar quantity
$X_{i_1 i_2 \dots i_m}$ made out of $m ( \leq N )$ position vectors
such that,
it picks up a minus sign under the exchange of any two indices and
vanishes whenever any two indices are same. The corresponding Jastrow
factor is
\be
J(\vec{r}_1, \vec{r}_2, \dots, \vec{r}_N)= \prod_{i_1 < i_2 < \dots < i_m}
(X_{i_1 i_2 \dots i_m} )^\lambda {\mid X_{i_1 i_2 \dots i_m} \mid}
^\alpha .
\label{eq1}
\ee
\noindent Under the exchange of `a'th and the`b'th particle, $J$ picks
up a factor $(-1)^{\lambda \delta}$ with $\delta$ given by,
\be
\delta = ^{N-2} C_{m-2} + 2 \ {\sum_{s=0}}
\ \ ^{a-1} C_p \ ^{b-a-1} C_q \ ^{N-b} C_r, \ p, q, r = 0, 1, \dots, m-2, \
b > a,
\label{eq2}
\ee  
\noindent where $^{N-2} C_{m-2}=\frac{(N-2)!}{(m-2)! (N-m)!}$ and
$s=p+2 q+r-m+2 $. The factor $^{N-2} C_{m-2}$ arises from those terms
where both the indices `a' and `b' are present. The second
term in the right hand side of (\ref{eq2}) receives contribution
from those terms having either the `a' or the `b' indices. The factor
two comes because the contribution is same for both the cases. Thus,
the second term is always even and $J$ picks up a minus
sign only for odd $\delta_1=^{N-2} C_{m-2}$. This immediately implies
that the fermionic nature of $J$ will be preserved for arbitrary $N$,
only when $m=2$.  One might also consider $X_{i_1 i_2 \dots i_m}$
to be (pseudo-)scalar/vector and try to construct $\tilde{J}$ by
taking linear combination of these \cite{cs2}, namely,
\be
\tilde{J}(\vec{r}_1, \vec{r}_2, \dots, \vec{r}_N) = \sum_{i_1 < i_2 <
\dots < i_m} a_{i_1 i_2 \dots i_m} X_{i_1 i_2 \dots i_m},
\label{eq3}
\ee
\noindent where $a_{i_1 i_2 \dots i_m}$ are the suitably
chosen coefficients. Unfortunately, $\tilde{J}$ has a definite
symmetry under the permutation of any two particles for
$N=m+1$ only.
Thus, the natural way to describe fermionic wave functions for
arbitrary number of particles is to construct Jastrow factor
(\ref{eq0}) with $X$ having only two indices.

The most general form of $X_{ij}$ in one dimension is $X_{ij}=
( x_i^k - x_j^k ) S_{ij}(x_i, x_j)$, where $k$ is a positive integer
and $S_{ij}$ is an
arbitrary function with the property that it is symmetric under the exchange of
`i' and `j'. Note that
any odd function with $(x_i^k - x_j^k) S_{ij}$ as its argument is also
a good candidate for $X_{ij}$. We will restrict ourselves throughout
this letter to the polynomial form of $X_{ij}$ only. The case
$k=1$ and $S_{ij}=1$ correspond to the Jastrow factor of the model
considered in Ref. \cite{cs}. The trivial
higher dimensional ( $D \geq 2$ ) generalization of this is
$X_{ij}= ( {\mid \vec{r}_i \mid}^\beta - {\mid \vec{r}_j \mid}^\beta)
S_{ij}(\vec{r}_i, \vec{r}_j)$, where $\beta$ is a real positive constant.
In $D \geq 3$ dimensions, this is the only
possibility satisfying all the requirements discussed above
as long as we are interested in constructing
$X_{ij}$ solely out of the position vectors. It might be noted at this
point that one can
construct many different $X_{ij}$ using the Cartesian components
of the position vectors. Most of these expressions for $X_{ij}$
can not be expressed
solely in terms of the position vectors and even if such a reduction
is possible for certain special cases, we will get back the expression
mentioned above. In other words, this is the unique choice for
(pseudo-)scalar Jastrow factor maintaining the basic criteria.
Since we are interested in constructing model
Hamiltonians, we consider  
$X_{ij}= {\mid \vec{r}_i \mid}^\beta - {\mid \vec{r}_j \mid}^\beta$
without loss of any generality. Now note that for $\beta=1$ and $\beta=2$,
the corresponding three-body interaction terms $W_3(\beta=1)$ and
$W_3(\beta=2)$ vanish, i.e., 
\be
W_3(\beta=1)= \frac{g^2}{2} 
\sum_{i \neq j \neq l} \frac{1}{({\mid \vec{r}_i \mid} -
{\mid \vec{r}_j \mid}) ({\mid \vec{r}_i \mid} - {\mid \vec{r}_l \mid})}
= 0,
\label{eq3.2}
\ee
\be
W_3(\beta=2) = 2 g^2 
\sum_{i \neq j \neq l} \frac{\vec{r}_l^2}{ ( \vec{r}_l^2 
- \vec{r}_j^2)( \vec{r}_l^2 - \vec{r}_i^2 )} = 0.
\label{eq3.3}
\ee
\noindent The identity (\ref{eq3.2}) is well known \cite{cs}.
To prove the identity (\ref{eq3.3}), first note that $W_3(\beta=2)$ can
be conveniently written as,
\be
W_3(\beta=2) = g^2 \sum_{i \neq j \neq l} \frac{ \vec{r}_i^2
+ \vec{r}_j^2 }{ ( \vec{r}_l^2 - \vec{r}_j^2)
( \vec{r}_l^2 - \vec{r}_i^2 )}.
\label{eq3.4}
\ee
\noindent Now exactly following the analysis of Ref. \cite{cs} for
the corresponding term in the CSM, it can be shown that $W_3(\beta=2)$
vanishes. 

We will construct below two different many-body Hamiltonians in
arbitrary dimensions, where these two Jastrow factors appear in the
corresponding wave functions. Before that, let us discuss on
$D=2$ dimensions, where there are more freedom in constructing $X_{ij}$.
In terms of the complex coordinates $z_i= x_i + i y_i$ and $\bar{z}_i=
x_i - i y_i$, the possible $X_{ij}$'s are
(i)$ X_{ij} = z_i^k - z_j^k$, (ii) $X_{ij} = {\mid z_i \mid}^\beta
- {\mid z_j \mid}^\beta$, and (iii)$ X_{ij}= z_i^k \bar{z}_j^k - 
z_j^k \bar{z}_i^k$. The expression (i) with $k=1$ appears in the
Laughlin's trial wave functions for the spin polarized electrons moving in
an external uniform magnetic field. No explicit model realizing such
correlations can be constructed since $\sum_i \partial_{z_i}
\partial_{\bar{z}_i} J(z_1, z_2, \dots, z_N) = 0$ for any $k$.
Note that we have freedom of multiplying the
right hand side of (i), (ii) and (iii) by $S_{ij}$.
In particular, one can
construct model Hamiltonian for $X_{ij}= (z_i^k-z_j^k) \ {\mid z_i
-z_j \mid}$. Unfortunately, the three-body interaction is non-vanishing
for such a choice. The expression (ii) has been discussed
above. The three-body interaction term is nonzero for the
third expression of $X_{ij}$, even though it offers novel correlations
\cite{mbs,pijush}.
   
Let us consider the Hamiltonian,
\be
H = - \frac{1}{2} \sum_{k=1}^N \vec{\bigtriangledown}_k^2
+ \frac{1}{2} \sum_{k=1}^N \vec{r}_k^2 + V_1(\beta)
+ V_2(\beta) + W_3(\beta), 
\label{eq4}
\ee
\noindent where $V_1(\beta)$, $V_2(\beta)$ and $W_3(\beta)$
are given by,
\bea
& & V_1(\beta)= \frac{\beta^2}{2} g (g-1)
\sum_{k \neq j} \frac{{\mid \vec{r}_k \mid}^{2(\beta-1)}}{(
{\mid \vec{r}_k \mid}^\beta -
{\mid \vec{r}_j \mid}^\beta)^2},\nonumber \\
& & V_2(\beta) =\frac{g \beta}{2} ( D + \beta -2) \sum_{k \neq j}
\frac{{\mid \vec{r}_k \mid}^{\beta-2}}{ {\mid \vec{r}_k \mid}^\beta
- {\mid \vec{r}_j \mid}^\beta},\nonumber \\ 
& & W_3(\beta) =  \frac{\beta^2}{2} g (g-1)
\sum_{i \neq j \neq k} \frac{{\mid \vec{r}_i \mid}^{2(\beta-1)}}{(
{\mid \vec{r}_i \mid}^\beta -
{\mid \vec{r}_j \mid}^\beta) (
{\mid \vec{r}_i \mid}^\beta -
{\mid \vec{r}_k \mid}^\beta) } .
\label{eq4p}
\eea
\noindent $g$ is a dimensionless coupling constant in (\ref{eq4p}).
We are working in the units $\bar{h}=m=w=1$, where $h=2 \pi \bar{h}$
is the Planck's constant, $m$ is the mass of each particle and
$w$ is the oscillator frequency.

The Hamiltonian
(\ref{eq4}) is defined in arbitrary $D$ dimensions. The one
dimensional results can be obtained by the replacement ${\mid
\vec{r}_i \mid} \rightarrow x_i$ and restricting $\beta$ to take
only positive integer values. The three-body
interaction term in the Hamiltonian vanishes for
$\beta=1, \ 2$ as mentioned previously. However, it is nonzero for
any other values of $\beta$.
Note that for $\beta=1$, $V_2(1)$ can be conveniently rewritten
as,
\be
V_2(1)=  - \frac{g}{4} (D-1) \sum_{k \neq j} \frac{1}{
{\mid \vec{r}_k \mid} {\mid \vec{r}_j \mid}} .
\label{eq4.1}
\ee
\noindent This term vanishes for $D=1$ and hence,
the Hamiltonian (\ref{eq4}) reduces to the one considered in \cite{cs}.
For $\beta=2$, $V_2(2)$ vanishes in arbitrary dimensions because it is
antisymmetric in `i' and `j'. The two-body interactions $V_1(2)$
can be alternatively written as,
\be
V_1(2) = 
\frac{g(g-1)}{2} \sum_{k \neq j} \left [ \frac{1}{({\mid \vec{r}_k
\mid} + {\mid \vec{r}_j \mid})^2}
+ \frac{1}{({\mid \vec{r}_k \mid} - {\mid \vec{r}_j \mid})^2} \right ] .
\label{eq4.2}
\ee
\noindent Note that this is related to the root structure of $D_N$
for $D=1$ \cite{pr}. However, for
$D \geq 2$, this interpretation is not valid since the interaction
is expressed in terms of the modulus of the position vectors. All
the results presented below are valid for arbitrary positive $\beta$.
However, we will mention only $\beta=1, \ 2$, while referring to the
Hamiltonians with only two-body interactions.

The Hamiltonian (\ref{eq4p}) can be written in terms of the
creation and anhilation operators as,
\be
H = E_0 + \frac{1}{2} \sum_i \vec{A}_{i}^\dagger . \vec{A}_i,
\label{eq6}
\ee 
\noindent where the ground state energy is $E_0= \frac{N D}{2} +
\frac{ g \beta}{2} N (N-1)$
and the anhilation operators are given by,
\be
\vec{A}_i = - i \vec{\bigtriangledown}_i - i \vec{r}_i + \beta g i
\sum_{i \neq j} \frac{{\mid \vec{r}_i \mid}^{\beta -
2} }{{\mid \vec{r}_i \mid}^\beta - {\mid \vec{r}_j \mid}^\beta} \vec{r}_i .
\label{eq7}
\ee
\noindent Note that the limit $g \rightarrow 0$ correctly reproduces
the ground state energy for
$N$ bosons confined in a harmonic oscillator potential. However, we do not
recover the corresponding result for $N$ fermions in $D \geq 2$ in the limit
$g \rightarrow 1$. In fact, $E_0$ with $g=1$ corresponds to
an excited state of $N$ fermions for $D \geq 2$. 
As we vary $g$ continuously for fixed $\beta$ and $D(\geq 2)$,
$E_0$ interpolates between the bosonic ground
state and an excited state of $N$ fermions.
This is a much more general problem in higher dimensional
many-body systems, like anyons in a harmonic oscillator potential
or those considered in Ref. \cite{cs2,mbs,pijush}, and at present
we do not have any solutions for this.  

The exact wave functions of (\ref{eq6}) are,
\be
\psi = \prod_{i < j} \left ( {\mid \vec{r}_i \mid}^\beta - 
{\mid \vec{r}_j \mid}^\beta \right )^\lambda \ {\mid \ 
{{\mid \vec{r}_i \mid}^\beta -
{\mid \vec{r}_j \mid}^\beta} \ \mid}^\alpha \
M(-n, E_0, t) \ exp(-\frac{1}{2}
\sum_k \vec{r}_k^2),
\label{eq10}
\ee
\noindent where $n$ is an integer, $t=\sum_i \vec{r}_i^2$ and $M(-n, E_0,t)$
is the confluent hypergeometric function. The wave
function vanishes whenever any two particles lie on the surface of
a $S^{D-1}$ whose center coincides with the origin of the coordinate
system. This automatically ensures that no two
particles are at the same position at the same time.
The energy eigen values corresponding to (\ref{eq10}) are $E_n=E_0 + 2 n$.
Note that this expression is similar to the CSM. Unfortunately, this is
not the complete spectrum. It seems that a further analysis similar to the
one developed in Ref. \cite{vok} in connection with the CSM is not possible
for this case. At present, we are unable to find the complete spectrum.

To conclude, we have constructed two different many-body Hamiltonians
in arbitrary dimensions with only two-body interactions. We obtained
the exact ground states as well as some of the excited states of these
Hamiltonians. We do not know of any physical system where these
model Hamiltonians are relevant. However, we expect that the exact
states of these models would serve as very good variational
wave-functions for physically interesting Hamiltonians of
nonrelativistic fermions. For example, the Jastrow factor,
\be
J(\vec{r}_1, \vec{r}_2, \dots, \vec{r}_N) =
\prod_{i<j} ({\mid \vec{r}_i \mid}^2
- {\mid \vec{r}_j \mid}^2 ) \ {\mid \vec{r}_i -
\vec{r}_j \mid}^{\gamma},
\label{last}
\ee
\noindent
is an exact eigen function of $N$ nonrelativistic 
fermions interacting with each other through $\frac{1}{{\mid
\vec{r}_i - \vec{r}_j \mid}^2}$ and some other physically
uninteresting three-body interactions. It is reasonable to
expect that a suitable choice of
the symmetric functions $S_{ij}$, which has to be found numerically in
general, multiplied with $J$ would eliminate
the effect of these unwanted many-body interactions.
Finally, we strongly believe that further analysis of
these two models along the line of CSM would provide us a better
understanding on the qualitative features of higher dimensional
many-body systems.
   
\acknowledgments I would like to thank Avinash Khare for a careful reading
of the manuscript and valuable comments.

\end{document}